\documentclass{aa}  

\usepackage{graphicx}
\usepackage{txfonts}
\usepackage{lipsum}
\usepackage{subcaption}
\usepackage{lscape}
\usepackage{placeins}

\usepackage{colortbl}
\usepackage{tabularx}
\usepackage{lipsum}
\usepackage{xcolor}
\usepackage{lineno}
\usepackage{caption}
\usepackage{hyperref}
\hypersetup{
  colorlinks,
  citecolor=blue,
  linkcolor=blue,
  urlcolor=blue}

\usepackage{natbib}
\bibpunct{(}{)}{;}{a}{}{,}

\newcommand{\target}{4U~1556--60}
\newcommand{\uJy}{$\mu$Jy}
\newcommand{\LRLX}{$\mathrm{L_{R}-L_{X}}$}
\newcommand{\ergps}{$\mathrm{erg~s^{-1}}$}

\newcommand{\ergpspcmsq}{$\mathrm{erg~s^{-1}~cm^{-2}}$}

\newcommand{\Gaia}{\textit{Gaia}}
\newcommand{\Ledd}{$L_{\mathrm{Edd}}$}
\newcommand{\Msun}{$\mathrm{M_{\odot}}$}
\newcommand{\degrees}{$^{\circ}$}

\begin{document}

\title{4U~1556--60 as a very faint neutron star X-ray binary at 700\,pc with an undetected radio jet}
\titlerunning{4U~1556--60 at 700~pc}

\author{
    E. C. Pattie\inst{1,2}\thanks{Corresponding author: e.c.pattie@uva.nl}
\and T. J. Maccarone\inst{2}
\and T. Russell\inst{3}
\and M. Bachetti\inst{4}
\and N. Degenaar\inst{1}
\and T. Kupfer\inst{5}
}

\institute{Anton Pannekoek Institute for Astronomy, University of Amsterdam, Postbus 94249, 1090 GE Amsterdam, The Netherlands
\and Department of Physics and Astronomy, Texas Tech University, Lubbock, TX 79409-1051, USA 
\and INAF, Istituto di Astrofisica Spaziale e Fisica Cosmica, Via U. La Malfa 153, I-90146 Palermo, Italy 
\and INAF Osservatorio Astronomicodi Cagliari, Viadella Scienza 5, 09047 Selargius (CA), Italy 
\and Hamburger Sternwarte, University of Hamburg, Gojenbergsweg 112, 21029 Hamburg, Germany
}

\date{Received MM DD, YYYY}

\abstract
   {\target\ is a low-mass X-ray binary that was discovered more than 50 years ago as a persistent X-ray source; however, very little was known about it. Recently, \Gaia\ obtained a parallax for the optical counterpart that places \target\ at a distance of only about 700~pc, making it one of the closest X-ray binaries known to date. This close distance drastically alters what was previously assumed about the source.}
   {We revisit \target\ in light of the newly determined distance of 700~pc, reinterpreting its literature and presenting new X-ray and radio observations to better understand various characteristics of the system.} 
   {We investigated the optical spectra and luminosity and the X-ray spectral and timing properties, and we performed the first targeted radio observation for the source in 45 years. These can be used to infer binary and accretion properties from independent methods.}
   {We conclude that a scenario in which \target\ is a candidate ultracompact neutron star X-ray binary at a distance of $\sim$700~pc is able to explain the observed properties of the source. It resides at a persistent X-ray luminosity of $\sim2\times10^{34}$~\ergps, an unusual value for a typical X-ray binary, but similar to several ultracompact systems. The ratio of the X-ray to optical luminosity is very high, also suggesting a physically small accretion disk. The radio jet is undetected with a very deep upper limit of $3\times10^{25}$~\ergps, which is about $10^3$ times fainter than the expected black hole jet correlation, strongly indicating a neutron star accretor. The X-ray spectrum is dominated by a power law, and the X-ray timing properties are also consistent with observations of other very low accretion rate X-ray binaries. No spin or orbital periodicity are found in the X-ray data. Future observations, especially to determine its orbital period, will further aid in understanding \target.}
   {}

\maketitle

\section{Introduction}

X-ray binaries (XRBs) are systems in which a neutron star (NS) or black hole (BH) accretes material from a companion star \citep{Done2007}. XRBs are usually transient, spending a significant amount of time in a faint, low activity quiescence between shorter, bright outburst events, though a small fraction are persistently accreting in an active state as well. There are several factors that govern the general behavior of an individual XRB, for example, the mass and spectral type of the companion, the orbital period, and whether the accretor is a black hole or neutron star.

\target\ was discovered as an X-ray source in 1972 by the \textit{Uhuru} mission \citep{Giacconi1972ApJ...178..281G} and was soon classified as an XRB with a low-mass donor star (LMXB) based on the identification of an optical counterpart \citep{Charles1979BAAS...11R.720C}. Further X-ray and optical studies were performed by \citet{Motch1989A&A...219..158M}, with an optical spectrum showing strong helium and weak hydrogen lines. The source distance was unknown and was assumed to be at a Galactic Center distance of $\sim$8\,kpc, resulting in an X-ray luminosity of $\sim 5 \times 10^{36}$~\ergps, which is a very typical X-ray luminosity for a persistently active LMXB \citep{XRBcats2023A&A...675A.199A}. The X-ray spectral shape was similar to XRBs with neutron star accretors; however, the lack of thermonuclear Type~I X-ray bursts in any of the X-ray data accumulated for \target\ at an assumed luminosity of $\sim$$10^{36}$~\ergps\ was problematic, though not strictly prohibitive, for a neutron star X-ray binary (NSXB) interpretation. There were also no X-ray pulsations detected, leaving the accretor nature undetermined. Later optical follow-up found strong Balmer lines in the optical spectrum and disfavored an ultracompact (i.e., a degenerate donor) interpretation for \target\ \citep{Nelemans2006MNRAS.370..255N}.

Recently, \Gaia\ Data Release 3 \citep{GaiaDR32023A&A...674A...1G} provided a parallax for the optical counterpart of \target\ of $1.4530\pm0.2644$~mas, corresponding to a distance of $688^{+153}_{-106}$\,pc. The median zero-point corrected geometric and photogeometric distances from \citet{Bailer-Jones2021AJ....161..147B} are 779\,pc and 760\,pc, respectively, and both are consistent with 700\,pc within 1$\sigma$. This \Gaia\ distance is an order of magnitude closer than previously assumed, which significantly changes many of the past interpretations of this source to date. 

We reevaluate the properties of \target\ based on a distance of $\sim$700\,pc, present new X-ray and radio data, and propose that \target\ is a very faint ultracompact NSXB with one of the weakest XRB radio jets observed to date. At a distance of about 700\,pc it is also one of the closest X-ray binaries known to date and the closest low-mass Roche lobe overflowing XRB \citep{Arnason2021MNRAS.502.5455A}.

\section{Observations and data reduction}

\subsection{X-ray data: NICER}

\target\ was observed with the Neutron Star Interior Composition ExploreR (NICER; \citealt{Gendreau2016SPIE.9905E..1HG}) for a total of 75\,ks over the date range of 2023 Apr 16 to 2023 May 04 (ObsIDs 658010101-658010110; a full list of observations are presented in Table~\ref{tab:NICER}). Data were processed in \texttt{HEASoft}\footnote{\citep{HEASOFT2014ascl.soft08004N}}  with \texttt{CALDB} version \texttt{20240206}. The data were reduced with the standard \texttt{nicerl2} pipeline calibration and filtering. The observations were analyzed individually as well as merged together with \texttt{niobsmerge}. Residual background flaring features persisted after calibration and these intervals were removed manually from the good time intervals with \texttt{xselect}. Spectra and light curves were then processed thorough \texttt{nicerl3-spect} and \texttt{nicerl3-lc} with \texttt{SCORPEON} background models, respectively\footnote{NICER data analysis threads containing \texttt{nicerl3} and \texttt{SCORPEON}: \url{https://heasarc.gsfc.nasa.gov/docs/nicer/analysis_threads/}}.

\subsection{Radio data: ATCA}
\target\ was observed with the Australia Telescope Compact Array (ATCA; \citealt{Wilson2011MNRAS.416..832W}) on 2024 Nov 09 to 2024 Nov 10 in a single 12 hour observation resulting in 10.3 hours on-source (Table~\ref{tab:ATCA}). The observation was recorded simultaneously in two frequency bands with 2\,GHz bandwidth each, centered on 5.5\,GHz and 9.0\,GHz. The array was in its most extended 6A configuration with a maximum baseline of 6\,km\footnote{\url{https://www.narrabri.atnf.csiro.au/operations/array_configurations/configurations.html}}. Flux density calibration was done using \texttt{PKS B1934-638}, and the phase calibrator was \texttt{B1613-586}. Data were processed using the Common Astronomy Software Applications (CASA; \citealt{CASA2022}) using a standard calibration process. Each of the two separated frequency bands ($4.5-6.5$\,GHz and $8-10$\,GHz) were calibrated independently, then the target data were imaged with the \texttt{tclean} task (\texttt{weighting=natural}), including phase self-calibration. The two self-calibrated frequency bands were then combined and imaged again with \texttt{tclean} and \texttt{gridder=mtmfs} to account for the large frequency span of the combined image.

\begin{table}
\caption{NICER X-ray observations of \target.}
\centering
    \setlength{\tabcolsep}{7pt}
	\begin{tabular}{ccc}
	    \hline\hline
	    ObsID & Date (YYYY MM DD) & Exposure (ks) \\
		\hline
		658010101 & 2023 04 16 & 1.4 \\
        658010102 & 2023 04 17 & 15.7 \\
        658010103 & 2023 04 18 & 14.2 \\
        658010104 & 2023 04 19 & 13.6 \\
        658010105 & 2023 04 20 & 1.4 \\
        658010106 & 2023 04 21 & 7.3 \\
        658010107 & 2023 04 22 & 9.0 \\
        658010108 & 2023 04 23 & 4.7 \\
        658010109 & 2023 04 27 & 6.3 \\
        658010110 & 2023 05 04 & 1.2 \\        
		\arrayrulecolor{black}\hline		
        Total &  & 74.9 \\
        \hline	\hline
	\end{tabular}
\label{tab:NICER}
\end{table}

\begin{table*}
\caption{ATCA radio observation of \target.}
\centering
    \setlength{\tabcolsep}{10pt}
	\begin{tabular}{ccccc}
	    \hline\hline
	    ObsID & Date (YYYY MM DD) & Obs. time & Freq. (GHz) & Upper limit ($3\sigma$, \uJy)\\
		\hline
		   &                          &       & 5.5 & 13.8 \\
        C3652 & 2024 11 09 -- 2024 11 10 & 12 hr & 9.0 & 17.3 \\
              &                          &       & 7.25 (combined) & 10.4 \\
        \hline\hline	
	\end{tabular}
	\tablefoot{Single multifrequency ATCA radio observation and nondetection results of \target.}
\label{tab:ATCA}
\end{table*}

\section{Results}

\subsection{Distance of 700~pc}

The \Gaia\ parallax for the optical counterpart of \target\ corresponds to a $1/\mathrm{parallax}$ distance of about 700\,pc. We first consider the scenario that this optical source is not the true counterpart and is rather an unrelated or interloper star. However, there are clearly broad optical emission lines in the spectrum of this counterpart that are attributed to accretion disks and not stellar emission, as well as a very blue continuum and an ultraviolet excess compared to a normal stellar spectral energy distribution \citep{Charles1979BAAS...11R.720C}. These features ensure that the emission of the optical counterpart source is indeed dominated by an accretion disk, strongly indicating its association with \target\ as an XRB, rather than being an unrelated superimposed star dominating the optical emission. Additionally, \target\ is located in a relatively low sky density region of the Galaxy ($l=+324^{\circ},\ b=-6^{\circ}$), where there is an average of 0.03 \Gaia\ sources per square arcsecond in the vicinity of \target. Thus the chance of an unresolved superposition of the \Gaia\ counterpart ($<0.4$" separation) based on the local stellar sky density is also low at $2.3\%$. 

We next consider that there may be a fainter interloper star that could affect the observed parallax shift, such that there is a superimposed interloper star that is much closer than 700~pc while \target\ is at a distance greater than 700~pc. This has indeed been observed to be the case for a different XRB, BW~Cir \citep{Gandhi2019}.  
However, the proper motion measurements from \Gaia\ Data Releases 2 and 3 for \target\ are consistent with each other within 1$\sigma$, unlike BW~Cir which showed a $2\sigma$ change in its proper motion between the two epochs as an artifact of the superposition evolving over time. 

A further restriction on a potential interloper star is the faintness of the optical and infrared counterparts of \target, and especially the latter. A very faint infrared counterpart was detected by the Vista Hemisphere Survey \citep[VHS;][]{McMahon2013Msngr.154...35M}, located 0.2" away from the \Gaia\ position at a J band magnitude of 18.4. 
Following the absolute J band magnitudes for M and L dwarf stars in \citet{Hawley2002AJ....123.3409H}, if an interloper star is located at, for example, $200$~pc it must be fainter than an L0 dwarf to not be brighter than the VHS counterpart. Then, using the inferred space density of L dwarfs from \citet{Cruz2007AJ....133..439C}, we estimate that the \Gaia\ superposition chance for L dwarfs out to 200~pc is $\sim10^{-5}$, thus it is extremely unlikely that there is an interloping faint nearby dwarf star.

A final consideration is that there may be an unknown cold white dwarf interloper in \Gaia, which may interfere with the optical parallax but would not distort the infrared spectral energy distribution of \target's counterpart (unlike a dwarf star above, which would be relatively bright in the infrared). \Gaia\ Data Release 3 has been examined for white dwarf stars by \citet{GentileFusillo2021MNRAS.508.3877G}, who find that the sky density near the Galactic Plane of white dwarfs with \textit{G} magnitudes $\leq 20$ is about 6 per square degree. Thus, the chance that there is a superimposed white dwarf is again extremely unlikely at $\sim 10^{-6}$.

Therefore, we argue that the optical source observed by \Gaia\ is indeed a singular and true counterpart to \target, providing a new distance of $\sim700$\,pc, with an optical \textit{G} magnitude of $19$ and an amplitude of variability of about half of a magnitude \citep{Motch1985SSRv...40..239M}. The distance estimates from \citet{Bailer-Jones2021AJ....161..147B}, the geometric value of $779^{+210}_{-153}$~pc and the photogeometric value of $760^{+198}_{-119}$~pc with 1$\sigma$ uncertainties, are consistent with a distance of $650-950$~pc. We subsequently use $700$~pc moving forward, as a slightly smaller or larger distance does not significantly change our conclusions for \target.

\subsection{X-ray data}

\begin{figure*}[ht!]
\sidecaption
\includegraphics[width=0.7\textwidth]{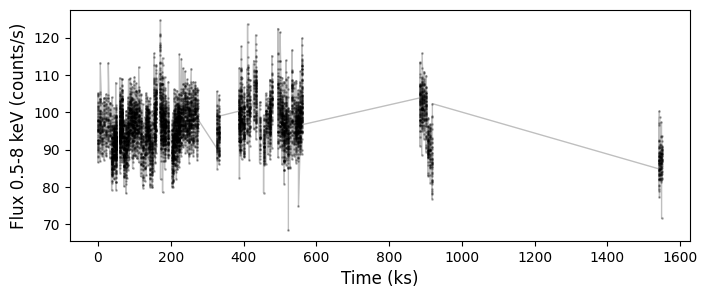}
    \captionsetup[figure]{justification=centering}
  \caption{NICER light curve of \target\ from 0.5-8\,keV with 8\,s cadence after excluding residual background flaring features. Flux errors are not plotted and are typically about 5\%. The stability of the flux observed in our NICER data justifies combining all observations to analyze spectral and timing properties. The light curve begins at MJD 60050.80902778.}
     \label{fig:lc_NICER}
\end{figure*}

Our recent NICER data are the first targeted X-ray observations of \target\ in about 15 years. With these data, we investigated the spectral components and the presence of an iron line, and searched for evidence of pulse and orbital periods. The X-ray flux over the 75\,ks of data is very stable ($91\%$ of light curve data are within $10\%$ of the median; Fig.~\ref{fig:lc_NICER}), therefore we combine all data for spectral analysis.

\subsubsection{NICER timing}

\begin{figure}
    \centering
    \includegraphics[width=0.45\textwidth]{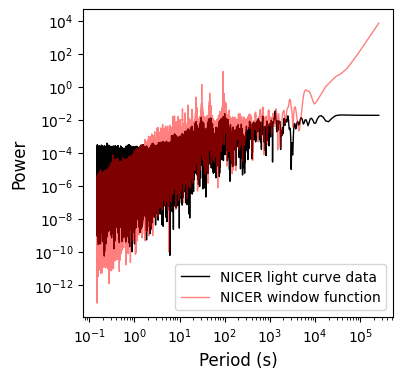}\\
    \caption{Lomb-Scargle of the full 75\,ks of NICER data with a light curve time resolution of 1\,s. There are no statistical features of the Lomb-Scargle periodogram that are not biased by the light curve sampling, as shown by the NICER window function Lomb-Scargle for comparison. Thus, the NICER data also do not show a sign of an orbital period.}
    \label{fig:lomb}
\end{figure}

The NICER data were searched for an orbital and pulse period in the energy range of $0.5-8$~keV. An orbital period was searched for by combining all NICER observations and extracting a 1\,s light curve with \texttt{nicerl3-lc}. A Lomb-Scargle periodogram (via \texttt{astropy}; \citealt{Astropy2013A&A...558A..33A}, \texttt{normalization = 'standard'}) was performed on both the NICER light curve and its window function to check for sampling biases. The resulting periodograms are shown in Fig.~\ref{fig:lomb}, with no statistical power at any frequency independent of the sampling biases indicated by the window function, thus there is no indication of an orbital period from the X-ray data. The sampling biases are consistent with the International Space Station's orbit of $93$ minutes, and harmonics of this value. The NICER data are sensitive to $\sim2\%$ sinusoidal modulations in the light curve, except near sampling-biased frequencies where the sensitivity worsens up to $\sim5\%$, for periods of up to $\sim1$~day. We also note that there were no Type~I X-ray bursts detected in the light curve, a continuing feature of this source from past X-ray data.

A pulse period was searched up to $1500$\,Hz in the combined $75$\,ks of NICER data using \texttt{stingray} \citep{stingray2019ApJ...881...39H} to construct the power spectral density (PSD) from the barycenter-corrected event file. There are no discernible features in the PSD, notably lacking a candidate spin period with a fractional pulse amplitude $5\times$RMS upper limit of $1.06\%$. The individual observations also displayed identically featureless PSDs. We did not perform an acceleration search, as a pulse period in an ultracompact system with a very low-mass donor would result in a relatively small smearing effect, and no candidate pulse period was identified. 
Potential orbital acceleration smearing of the pulse period for a very low-mass donor star (discussed further later) is small ($\lesssim0.3$\,Hz for a 500\,Hz pulsar, which would have been resolved), thus we are very sensitive to a potential spin period without an acceleration search.

\subsubsection{NICER spectrum}

\begin{figure}[ht]
    \centering
    \includegraphics[width=0.45\textwidth]{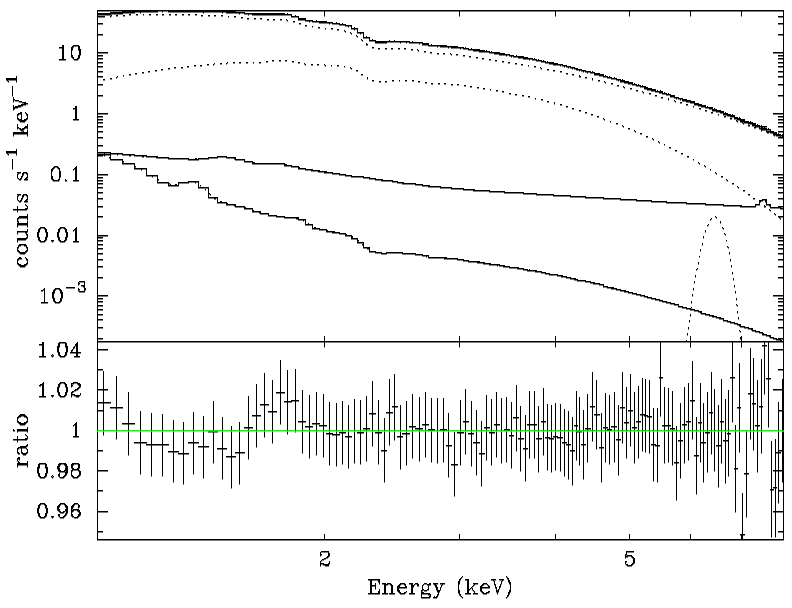}\\
    \caption{X-ray spectrum of \target\ from NICER from all observations combined. The spectrum is fit with a \texttt{powerlaw} + \texttt{bbodyrad}, and a \texttt{gaussian} for the iron line. Background models are also shown. Full parameters are in Table~\ref{tab:NICERspect}.}
    \label{fig:xspec}
\end{figure}

The spectrum of the combined $75$\,ks of \target\ (Fig.~\ref{fig:xspec}) (\texttt{xspec} version \texttt{12.14.0h}; \citealt{xspec1996ASPC..101...17A}) is consistent with a nonthermal-dominated hard spectral state, with a flux of $\sim$$3 \times 10^{-10}$~\ergpspcmsq, corresponding to a $2-10$~keV luminosity of $\sim$$2 \times 10^{34}$~\ergps\ at a distance of 700\,pc. The spectrum is fit with a \texttt{powerlaw + bbodyrad}, and a \texttt{gaussian} for the iron line with an equivalent width of $\sim$8~eV ($\chi^2/\mathrm{dof} = 41.37/115$). Abundances and photoelectric cross sections for \texttt{phabs} are from \texttt{angr} \citep{Anders1989GeCoA..53..197A} and \texttt{vern} \citep{Verner1996ApJ...465..487V}, respectively. Full fit parameters are in Table~\ref{tab:NICERspect}. The data were also similarly fit well statistically with a \texttt{diskbb} instead of a \texttt{bbodyrad} ($\chi^2/\mathrm{dof} = 37.90/115$); however, the normalization for \texttt{diskbb} in this case resulted in an apparent inner disk radius of $<1$~km at $700$~pc, therefore we deemed \texttt{diskbb} unphysical and instead favor the \texttt{powerlaw + bbodyrad} to fit the continuum. The \texttt{bbodyrad} with a temperature of $\sim$$0.9$~keV has a characteristic radius of $0.21\pm0.02$~km from its normalization parameter.

We tested additional model scenarios for the spectrum. Fixing the \texttt{bbodyrad} component to a larger radius of $5$~km resulted in a poor fit with $\chi^2/\mathrm{dof} = 559.61/115$ with strongly structured residuals. Removing the blackbody component and fitting the continuum with only a \texttt{powerlaw} resulted in a similarly poor fit with $\chi^2/\mathrm{dof} = 618.12/116$, again with strongly structured residuals. An attempt to use the \texttt{nsatmos} model \citep{Heinke2006ApJ...644.1090H} instead of \texttt{bbodyrad} resulted in a failure to fit the spectrum due to the constrained range of the temperature parameter. A \texttt{comptt}-only model \citep{Titarchuk1994ApJ...434..570T} was also poorly fit with $\chi^2/\mathrm{dof} = 305/117$. The spectrum was fit equally well with \texttt{comptt + bbodyrad} with $\chi^2/\mathrm{dof} = 50.54/115$ and blackbody parameters consistent with the \texttt{powerlaw + bbodyrad} of a hot and small ($1.05$~keV, $5.6$ normalization) thermal component. Thus, the spectral fit necessitates a physically very small and hot thermal component in addition to a dominating nonthermal component.

\begin{table}
\caption{NICER X-ray spectral fit of \target.}
\centering
    \setlength{\tabcolsep}{7pt}
	\begin{tabular}{ccc}
	    \hline\hline
	    Model & Parameter & Value \\
		\hline
		phabs & nH ($\times10^{22}$) & $0.34^{+0.016}_{-0.015}$\\
        \arrayrulecolor{lightgray}\hline
        bbodyrad & kT (keV)& $0.92^{+0.03}_{-0.04}$ \\
              & norm & $9.12^{+1.75}_{-1.56}$ \\
              & {\it $L_X~(erg~s^{-1})$} & {\it $\sim\mathit{4\times 10^{33}}$} \\
              & {\it $radius~(km)$} & {\it $\mathit{0.21\pm0.02}$} \\
        \hline
        powerlaw & PhoIndex & $1.53^{+0.04}_{-0.04}$ \\
                 & norm & $5.66^{+0.4}_{-0.4} \times 10^{-2}$ \\
        \hline
        gaussian & LineE (keV) & $6.46^{+0.14}_{-0.17}$ \\
                 & Sigma (keV) & $0.16^{+0.17}_{-0.16}$ \\
                 & norm & $2.83^{+0.002}_{-0.002} \times 10^{-5}$ \\  
                 & {\it EqW (eV)} & {\it $\mathit{7.83}$} \\
		\arrayrulecolor{black}\hline		
        Chi-squared/dof & 41.37/115 & \\
        Flux (2--10 keV) & (\ergpspcmsq) & $3.46 \times 10^{-10}$ \\
        $\mathrm{L_{X}}$ (2--10 keV) & (\ergps) & $1.97 \times 10^{34}$ \\
        \hline\hline
	\end{tabular}
	\tablefoot{Luminosities assume a distance of 700\,pc. Data were fit in \texttt{xspec}.}
\label{tab:NICERspect}
\end{table}

\subsubsection{Other X-ray data}

\target\ is monitored by the Monitor of All Sky X-ray Image (MAXI; \citealt{MAXI2009PASJ...61..999M}) telescope (source name 1H~1556--605), and over the last $\sim$15 years its X-ray light curve has not shown any significant long-term variability, reinforcing the view that \target\ persistently accretes at a very stable rate, corresponding to a median $2-8$\,keV luminosity value of $3 \times 10^{34}$~\ergps. Other targeted X-ray data in the literature over the decades have also found similar fluxes and spectral shapes of a power law + thermal component \citep{Motch1989A&A...219..158M, Farinelli2003ASPC..308..283F}. There have been no outbursts or Type~I bursts detected for \target.

\subsection{Radio data}

\begin{figure}
    \centering
    \includegraphics[width=0.45\textwidth]{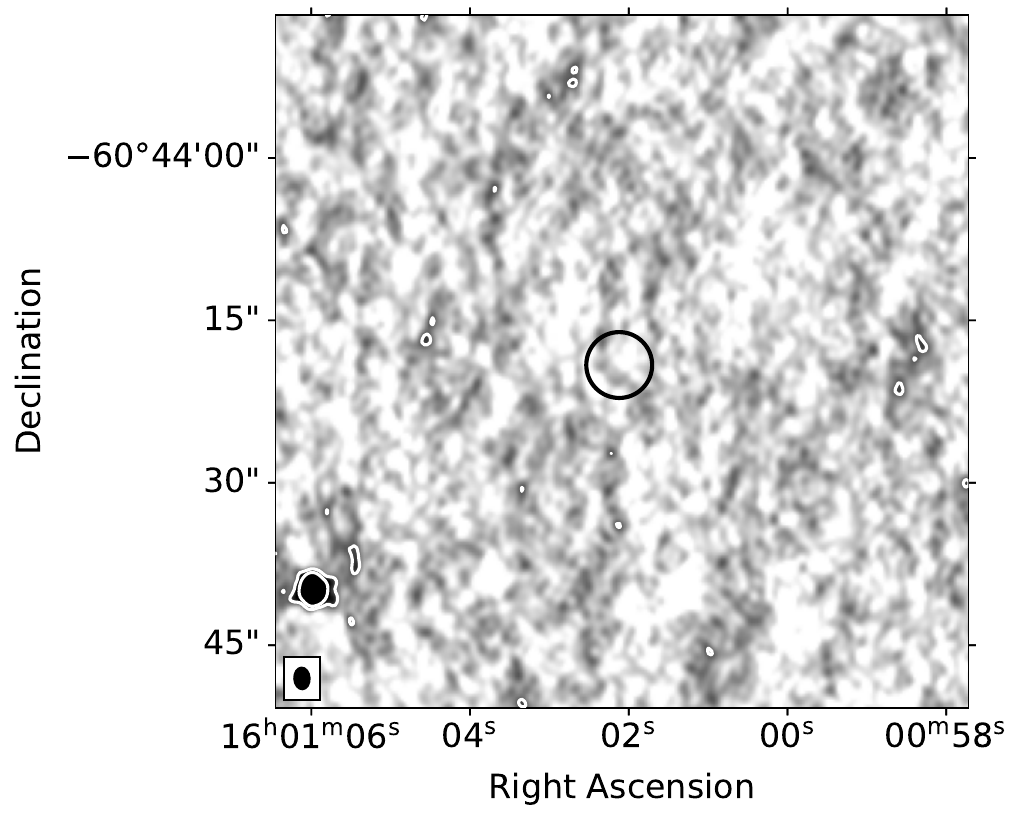}\\
    \caption{ATCA radio image of \target\ at the combined frequency of 7.25\,GHz. The black circle is centered on the {\it{Gaia}} optical counterpart for visual reference. The contours plotted are $3\sigma$ and $10\sigma$, where $\sigma$ is the background RMS value of $3.5\,\mu$Jy. The bright source in the lower left corner has a peak flux density of 174\uJy\ ($\sim50\sigma)$. }
    \label{fig:atca}
\end{figure}

The ATCA radio data did not detect a radio counterpart to \target\ with a $3\sigma$ upper limit of $10.4$~\uJy\ at 7.25~GHz (Fig.~\ref{fig:atca}), corresponding to an upper limit for the radio luminosity of $\sim$$3 \times 10^{25}$~\ergps\ at 5\,GHz assuming a flat spectral index, as expected for LMXBs in their hard states \citep{Fender2001MNRAS.322...31F}. This result and other XRB X-ray and radio luminosities are shown in Fig.~\ref{fig:lrlx}.

\section{Discussion}

\subsection{A neutron star accretor}

\begin{figure*}[ht!]
\sidecaption
\includegraphics[width=0.6\textwidth]{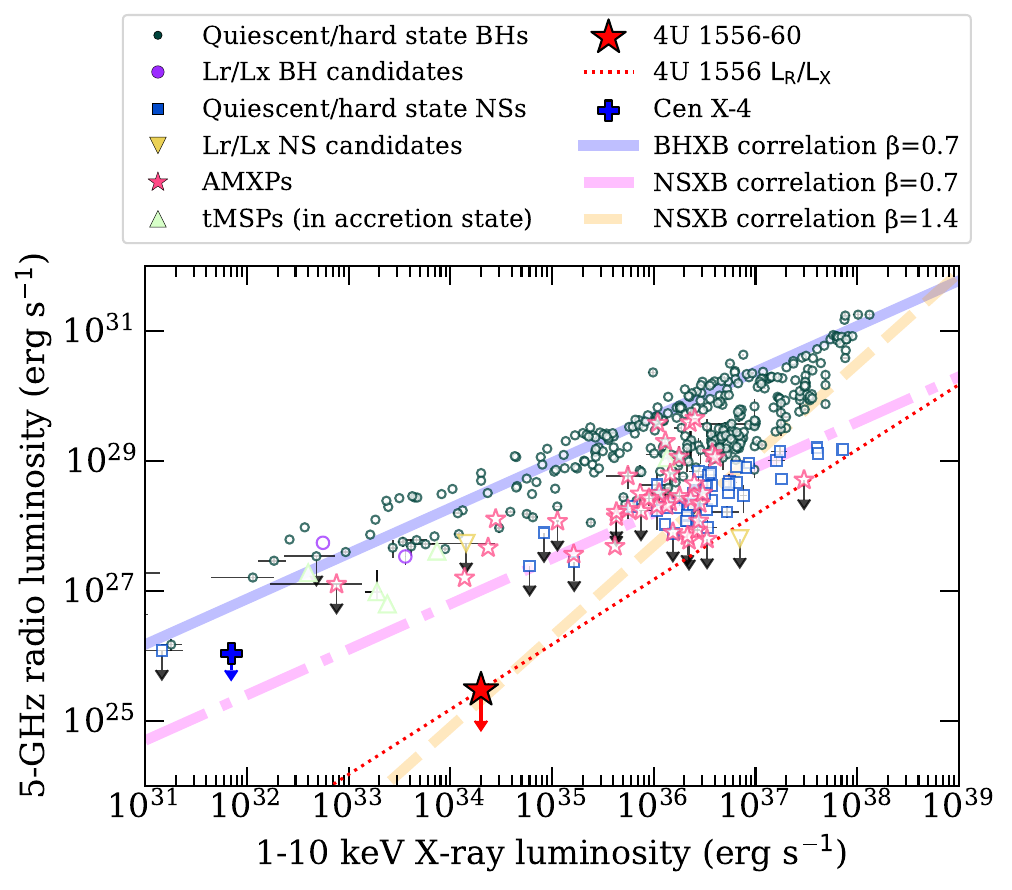}
    \captionsetup[figure]{justification=centering}
  \caption{Plot of X-ray binary radio and X-ray luminosities from \citet{arash_bahramian_2022_7059313}, with our radio nondetection of \target\ indicated with a red star and a line of constant $\mathrm{L_R/L_X}$ plotted for reference at other source distances. Three correlation lines are plotted for visual reference: solid blue for BHXBs; dash-dot magenta for NSXBs with $\beta=0.7$; and dashed orange for NSXBs with $\beta=1.4$. AMXPs are accreting millisecond pulsars with stronger magnetic fields than general NSXBs, and tMSPs are transitional millisecond pulsars whose radio emission during their XRB-like state is thought to not arise from a jet alone. The very low upper limit of \target's radio emission for its X-ray luminosity is strongly indicative of a neutron star accretor, as it is about $1000$ times fainter than expected for a black hole accretor. The previous deepest upper limit in radio luminosity for a NSXB, Cen~X--4 in quiescence at $\sim1.2$~kpc \citep{vandenEijnden2022MNRAS.516.2641V}, is also shown as a blue plus sign.}
     \label{fig:lrlx}
\end{figure*}

It was generally believed that the accretor in \target\ was a neutron star, primarily based on its X-ray spectral properties. Our recent radio data, an upper limit of $3\times10^{25}$~\ergps\ and the first targeted observation since \citet{Duldig1979MNRAS.187..567D} ($<5$~mJy nondetection), provide the strongest evidence that the accretor in \target\ is indeed a neutron star. The correlation between radio and X-ray luminosity \citep{Fender2004, Gallo2018MNRAS.478L.132G}, \LRLX, shown in Fig.~\ref{fig:lrlx}, can be used as a diagnostic of the nature of the accretor for unambiguous cases due to the ubiquitous presence of black hole X-ray binary (BHXB) jets in hard spectral states, including detections in quiescence for some brighter systems \citep[e.g.,][]{Gallo2006, Corbel2008MNRAS.389.1697C, Soleri2011MNRAS.413.2269S}. The upper limit we obtained shows that \target\ has a radio jet that is at least three orders of magnitude fainter than the BHXB population at a similar X-ray luminosity. Thus, this upper limit on the radio jet luminosity indicates that \target\ almost certainly hosts a neutron star that is producing a much weaker jet relative to the expectation of BHXBs at similar X-ray luminosities. We return to further implications of \target's radio jet luminosity later in the discussion in Section~\ref{sec:radio_jet}.

\subsubsection{Lack of Type~I bursts}

\target\ has never shown Type~I bursts, which are a very common though not universal feature of accreting neutron stars with low magnetic field strengths typical of NS LMXBs \citep{Lewin1993SSRv...62..223L, Degenaar2018SSRv..214...15D, Galloway2020ApJS..249...32G}. We can calculate the expected burst rate for hydrogen bursts using $\alpha = L_{\mathrm{pers}} \times t_{\mathrm{rec}} / E_{\mathrm{burst}}$, where $\alpha$ is a value between 40-100 \citep{Strohmayer2006csxs.book..113S}, $L_{\mathrm{pers}}$ is the persistent X-ray luminosity, $t_{\mathrm{rec}}$ is the burst recurrence time, and $E_{\mathrm{burst}}$ is the fluence of the burst. We estimate that the recurrence time of standard hydrogen bursts for \target\ is $\gtrsim10^3$\,ks, or about 10 bursts per year, assuming typical LMXB accretion disk abundances dominated by hydrogen. With this inferred burst rate, the lack of bursts detected in the accumulated but sporadic X-ray data of \target\ to date is indeed expected. 

This is still a conservative under-estimation of the recurrence time due to the very low X-ray luminosity of \target, as the neutron star itself will be much cooler than more typical, higher persistent X-ray luminosities thus requiring a more massive buildup of material before ignition \citep{Peng2007AIPC..924..513P}. This may result instead in even more infrequent ``intermediate" duration bursts which have been observed in other NSXBs at similarly low X-ray luminosities \citep[e.g.,][]{Degenaar2010MNRAS.404.1591D, Alizai2023MNRAS.521.3608A}. Thus, \target\ may be slowly building up a very thick layer of hydrogen or helium, enabled by the low accretion rate delaying the fuel reaching the critical ignition temperature, and possibly steadily fusing hydrogen into helium as well. It would then very rarely exhibit a luminous intermediate duration burst with recurrence times possibly as long as several years or more \citep{Kuulkers2009A&A...503..889K, Alizai2023MNRAS.521.3608A}. If \target\ does burst, then it may be a nearby analog or member of the ``burst-only" NSXBs usually observed at Galactic Center distances where low luminosity persistent X-ray emission is often not detected \citep[e.g.,][]{Cocchi2001A&A...378L..37C, Cornelisse2002A&A...392..931C}. 

There are no reported \textit{Swift} Burst Alert Telescope (BAT; \citealt{SwiftBAT2005SSRv..120..143B}) triggers in the direction of \target\ (none within $1^\circ$), and similarly there is no over-density of \textit{Fermi} Gamma-Ray Burst Monitor (GBM; \citealt{FermiGBM2009ApJ...702..791M}) triggers. At 700~pc, BAT and GBM are able to detect Type~I bursts as they have for other NSXBs at greater distances \citep{Linares2012ApJ...760..133L, Jenke2018yCat..18260228J, intZand2019A&A...621A..53I, Lin2020ApJ...903...37L}.
Thus, BAT and GBM both provide additional indications that the burst rate of \target\ is extremely low, of order $\lesssim1$~burst/year.

Another potential is that \target\ hosts a very strong magnetic field of $\gtrsim10^{12}$~G, which observationally inhibits the ability of NSXBs to burst. This is believed to be due to magnetic confinement of the accreting material driving persistent burning \citep{Joss1980ApJ...238..287J}. This possibility will be discussed further below in Section~\ref{sec:xray}.

\subsection{Ultracompact nature and binary parameters}

Ultracompact X-ray binaries (UCXBs) are a class of XRBs that have short orbital periods of $\lesssim80$~minutes \citep{Savonije1986A&A...155...51S, vanHaaften2012A&A...537A.104V}. Main sequence donor stars cannot fit within their Roche lobe at these orbital periods, thus UCXB donors are necessarily degenerate white dwarfs or semi-degenerate helium stars. UCXBs are of particular interest, for example, as products of common envelope evolution \citep{Paczynski1976IAUS...73...75P, Ivanova2008MNRAS.386..553I}, as progenitors of radio millisecond pulsars \citep{Tauris2012MNRAS.425.1601T}, and as gravitational wave sources \citep{Nelemans2001A&A...375..890N, Chen2020ApJ...900L...8C}. They are hydrogen-poor and thus some other their behaviors may be affected, such as bursts in NS UCXBs. There are about 50 known or candidate UCXBs to date \citep{ArmasPadilla2023A&A...677A.186A}, with 20 of them confirmed via orbital period measurements. Their population disproportionally resides in globular clusters likely due to the much higher rate of dynamical interactions between stars there \citep{Verbunt1987ApJ...312L..23V, Ivanova2008MNRAS.386..553I}.

We present multiple methods to estimate or explain the orbital period in \target. Although none of these individually are robust enough to determine the orbital period definitively, altogether these suggest that the orbital period is likely to be ultracompact.

\subsubsection{Optical and infrared luminosity}
\label{sec:OIR_luminosity}

The optical magnitude of the \Gaia\ counterpart is about 19, with a distance modulus of 9.2 magnitudes and an additional 1 magnitude to account for extinction based on the NASA/IPAC Extragalactic Database calculator\footnote{\url{https://ned.ipac.caltech.edu/extinction_calculator}}. Thus, we estimate that the absolute optical magnitude of \target\ is about 8.8 magnitudes. The VHS infrared J band counterpart is about 18.4 magnitudes, thus an absolute magnitude of 9.4.

In XRBs, the observed optical and infrared (OIR) emission is a combination of the donor star's light and reprocessed X-ray emission from the accretion disk, and possibly jet emission. In LMXBs, the disk reprocessing dominates the optical emission when the source is not in quiescence (\citealt{vanParadijs1994A&A...290..133V}; hereafter vPM94). The optical spectra of \target\ did not show any stellar features \citep{Motch1989A&A...219..158M, Nelemans2006MNRAS.370..255N}, thus we can assume that the optical emission is similarly dominated by the accretion disk rather than by the donor star. We do not consider OIR jet emission due to the radio nondetection.

Simple test cases for the orbital period can be performed given the requirement that \target\ must have a Roche lobe filling donor star in an orbit with an assumed $1.4$~\Msun\ neutron star. If 10\% of the optical light is emitted from the companion (and the rest from the disk reprocessing X-ray emission), a donor star with an absolute visible magnitude of 11.3 corresponds to a type M3, and thus an orbital period of about 3 hours. If only 1\% of the optical light is from the companion, then the 13.8 magnitude M5 donor star would have an orbital period of 2 hours.

Similarly using the VHS counterpart, a star with an absolute J band magnitude of 9.4 would correspond to an early L dwarf. Assuming this L dwarf produces all of the observed infrared counterpart's emission, the orbital period would be only 2 hours, and again this is an over-estimation of the donor emission contribution and thus orbital period. Therefore, the faintness of the OIR counterpart alone places strong upper limits on the orbital period of \target\ due to the requirement that the donor star fill its Roche lobe in this system.

\subsubsection{X-ray:OIR ratio}
\label{sec:xray_optical_ratio}

The X-ray luminosity can also be utilized in combination with OIR to estimate the size of the accretion disk, and therefore estimate the orbital period of the binary. A very high X-ray to optical luminosity ratio indicates that there is a physically small accretion disk surface area which is reprocessing the X-ray emission. vPM94 utilizes this to roughly estimate the orbital period of LMXBs based on their X-ray luminosity and absolute optical magnitude during times of active accretion when the donor star emission is a negligible contribution to the total optical light. 

Following this for \target\ with $L/L_{\mathrm{Edd}}\approx 10^{-4}$ and $M_V$ of $8.8$, the orbital period is estimated to be $60^{+110}_{-60}$\,s. Though an orbital period of $60$\,s is certainly unphysical, the error of this is also quite large which we find is due to the quoted uncertainties of the best fit slope equation in vPM94, combined with the overall low X-ray and optical luminosities for \target. There is also significant scatter in the original relationship derived by vPM94 of about $0.5$ in $log~\Sigma$, which can increase the inferred period to be $6^{+10}_{-6}$\,minutes at $+0.5~log~\Sigma$, or $30^{+54}_{-30}$~minutes at $+1~log~\Sigma$.

\begin{figure}
    \centering
    \includegraphics[width=0.45\textwidth]{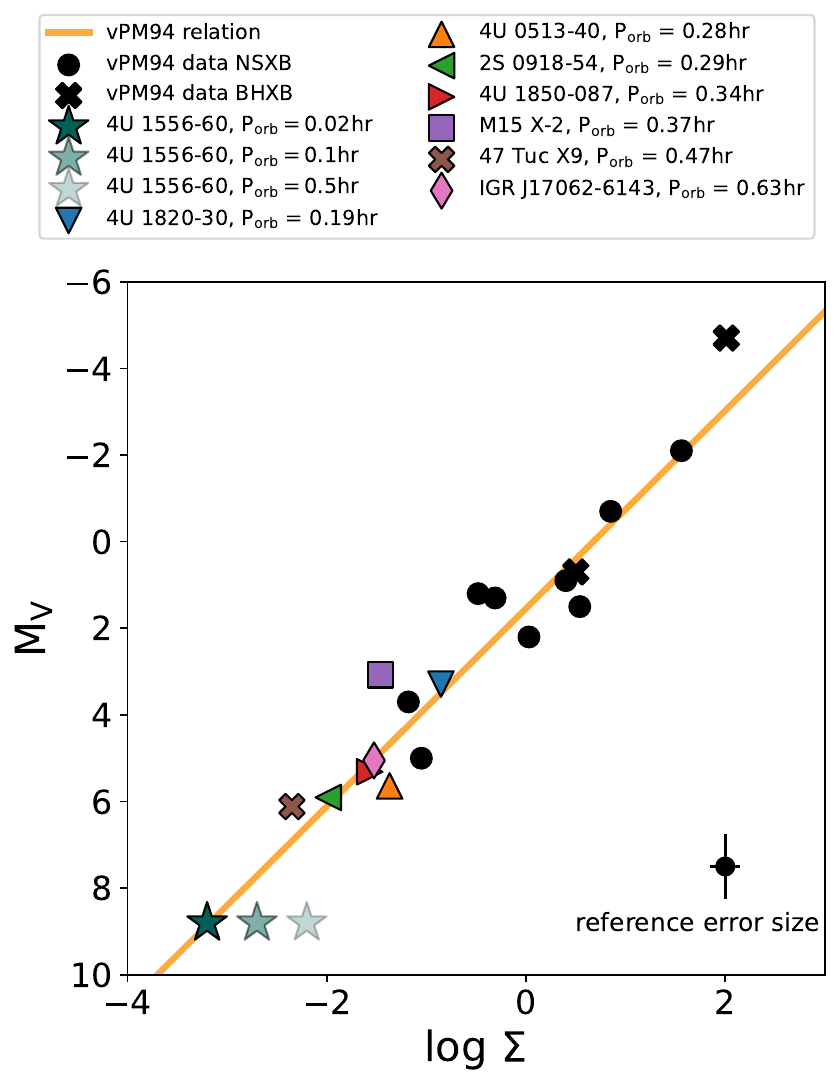}\\
    \caption{Reproduced plot relating the absolute visual magnitude of LMXBs with the parameter $\Sigma = (L_X/L_{Edd})^{1/2} (P/1\,hr)^{2/3}$ from \citet{vanParadijs1994A&A...290..133V}. vPM94 NSXBs are black circles and BHXBs are black X's, with reference errors on a point in the lower right. We have added the nominal vPM94 relation line in orange. We have also added a sample of known UCXBs from \citet{ArmasPadilla2023A&A...677A.186A} in various colors and marker shapes. Estimations for the orbital period of \target\ are shown in three blue stars with decreasing opacity, at $+0$, $+0.5$, and $+1$~$\mathrm{log~\Sigma}$ respectively.}
    \label{fig:vanPar94}
\end{figure}

These results are plotted in Fig.~\ref{fig:vanPar94}, including the original data as well as a sample of UCXBs from \citet{ArmasPadilla2023A&A...677A.186A} which have X-ray and optical fluxes and known orbital periods. There is indeed scatter in the relation, in part from inclination angle effects, varying mass ratios, as well as non-simultaneity of the multiwavelength data. 
Overall, we believe this to broadly indicate that \target\ indeed has a very short orbital period as evidenced by its faint optical counterpart, even if the above relation is not as precise at low X-ray and optical luminosities. 

Similar to vPM94, \citet{Russell2006MNRAS.371.1334R} investigated the correlation between X-ray and OIR emission of XRBs, considering both disk and jet contributions to the OIR emission and fitting BH and NS XRBs separately. Assuming that there is negligible OIR jet emission in \target\ due to the radio nondetection, the \Gaia\ counterpart again corresponds nominally to an unreasonably short orbital period of $\sim4$~minutes for a NSXB, but considering the scatter around this relation the inferred orbital period can reasonably be extended to 40~minutes. Using the VHS counterpart at $2.14~\mu$m instead returned similar results with a nominally very short orbital period.

\subsubsection{Unusual persistent X-ray luminosity}

The X-ray luminosity of \target, stable since its discovery \citep[e.g.,][and since 2009 from MAXI]{Wood1984ApJS...56..507W, Motch1989A&A...219..158M, Farinelli2003A&A...402.1021F}, is at an unusual value of about $10^{34}$~\ergps. It is significantly brighter than typical quiescent XRBs ($\sim10^{32}$~\ergps, e.g., \citealt{Heinke2003ApJ...598..501H, Wijnands2017JApA...38...49W}), while also being much fainter than the majority of other persistently active systems ($\gtrsim 10^{36}$~\ergps, e.g., \citealt{intZand2007A&A...465..953I}). 
However, UCXBs are able keep their small accretion disks ionized at low X-ray luminosities via irradiation from the inner regions \citep{vanParadijs1996ApJ...464L.139V, Dubus1999MNRAS.303..139D}, thus allowing them to be persistently active at lower X-ray luminosities than the more common longer period systems \citep{Lasota2001NewAR..45..449L}.

This persistent luminosity may also be related to a high mass transfer rate from the donor star. Transient XRBs have lower mass accretion than transfer rates in quiescence, thus causing the accretion disk mass to build up and trigger an instability. The subsequent outburst accretion rate exceeds the mass transfer rate and depletes the disk, ending the outburst \citep{Lasota2001NewAR..45..449L}. Persistent systems may be accreting at close to the mass transfer rate, thus preventing both the build-up and depletion of mass in the disk. It may also be that a high mass transfer rate keeps the outer disk ionized, hence driving the observed persistent accretion rate over long periods of time \citep{Lasota2001NewAR..45..449L, intZand2007A&A...465..953I}. This accretion rate would be lower than in most transient systems during outburst but also higher than typical quiescent ones. This is especially relevant for UCXBs where the mass transfer rate is driven by gravitational wave radiation \citep{vanHaaften2012A&A...537A.104V}, resulting in high mass transfer rates for a low-mass donor star overflowing its Roche lobe.

Following \citet{vanHaaften2012A&A...537A.104V} for ultracompact systems, and assuming that the $2-10$~keV X-ray luminosity is half of the bolometric luminosity of \target\ \citep{Anastasopoulou2022MNRAS.513.1400A}, we estimate that the mass transfer rate from the donor to the disk is $2 \times 10^{-11}$~\Msun/year ($10^{15}$~g/s) for a neutron star accretor. This mass transfer rate for a white dwarf donor driven by gravitational wave radiation corresponds to an orbital period of $20-30$ minutes (Fig.~2 in \citealt{vanHaaften2012A&A...537A.104V}), and again points toward an ultracompact nature for \target.

\target's X-ray luminosity is indeed similar to a 28~minute ultracompact candidate BHXB that appears to be in an unusually bright quiescence of $\mathrm{L_X \gtrsim 10^{33}}$~\ergps, 47~Tuc~X9 \citep{Bahramian2017MNRAS.467.2199B}. This persistent luminosity is believed to be driven by an elevated mass transfer rate from the degenerate donor in quiescence. 
A similar scenario is theorized for the persistent NSXB RX~J1718.4--4029 \citep{intZand2009A&A...506..857I}, which has an X-ray luminosity of $\sim8\times10^{34}$~\ergps, though an orbital period has not been confirmed in this system. 
The confirmed NS UCXB IGR~J17062--6143 \citep{Strohmayer2018ApJ...858L..13S} was discovered during an outburst in 2006 \citep{Churazov2007A&A...467..529C}, and has since resided at a persistent luminosity of $\sim10^{35}$~\ergps, a particularly high non-outburst value for a transient system. 

Systems such as these with $L_X$ of $10^{34-35}$~\ergps\ are known as very faint X-ray binaries (VFXBs) \citep[e.g.,][]{Pavlinsky1994ApJ...425..110P, Heinke2015MNRAS.447.3034H}. Many VFXBs are also UCXBs, though giant donor stars have also been identified in two NS VFXBs \citep{Shaw2020MNRAS.492.4344S, Shaw2024MNRAS.527.7603S}. However, the faintness of the OIR counterpart effectively rules out a giant donor scenario for \target\ at any Galactic distance.

\subsubsection{Presence of hydrogen}

The presence of hydrogen in one optical spectrum is notable given that \target\ is estimated above to be ultracompact, limiting the potential orbital period and evolution history. The apparently variable strength of hydrogen lines (very weak but present in \citealt{Motch1989A&A...219..158M}, and strongly found later in \citealt{Nelemans2006MNRAS.370..255N}) could indicate that there is a small amount of hydrogen in the accretion disk, such that small fluctuations in ionization lead to large changes in the strength of the Balmer lines. Thus, an ultracompact system with trace amounts of hydrogen in the donor star is a possibility for \target. 

Hydrogen is very problematic for UCXBs that have very short orbital periods of $\lesssim20$~minutes \citep[e.g.,][]{Podsiadlowski2003MNRAS.340.1214P, Belloni2023A&A...678A..34B}, but is possible for slightly longer ones. Based on simulations of white dwarf accretors, it is indeed possible for a small fraction of hydrogen to remain in the donor envelope at short orbital periods if the donor is an evolved star rather than a white dwarf \citep{Nelemans2010MNRAS.401.1347N, Goliasch2015ApJ...809...80G, Kalomeni2016ApJ...833...83K}. This scenario is similar to the $51$~minute accreting white dwarf system ZTF~J1813+4251 \citep{Burdge2022Natur.610..467B}, which has observed hydrogen lines and is expected to reach a minimum orbital period of $\sim20$~minutes. Thus, it is perhaps more likely that \target\ has an orbital period that is greater than $20$~minutes, such that the donor is an evolved star that has not yet been fully stripped of hydrogen, and the binary is currently evolving toward a period minimum. In this case, \target\ would likely be the product of two common envelope phases. A UCXB with an evolved donor should be a rare occurrence; if this scenario is indeed true, \target\ being the only known member of this class is unsurprising given that there are only about 50 known or candidate UCXBs to date.

\subsubsection{Inclination angle}

No eclipsing or dipping features in the X-ray light curve have been observed, indicating that \target\ is not highly inclined. A further estimation of its inclination angle can be performed by utilizing its optical spectrum. We take the full width at half maximum (FWHM) of the strong hydrogen line in \citet{Nelemans2006MNRAS.370..255N} of $\mathrm{540~km~s^{-1}}$ and use the relation of \citet{Casares2015ApJ...808...80C} between the FWHM of H$\alpha$ and K2 (the orbital velocity of the donor star) to estimate the required inclination angle given a constraint on the orbital period. For an orbital period of 30~minutes, $i$ is 8\degrees, for 1~hour it is 11\degrees, and for a much longer period of 6 hours it is still only 20\degrees (nominal errors of $\pm3$\degrees). Thus, \target\ is likely a low inclination system for any reasonable orbital period based on its observed narrow H$\alpha$ line.

\subsubsection{A mHz gravitational wave source}

Lastly, we briefly consider that a UCXB at a distance of 700\,pc is a very appealing prospect for mHz gravitational wave (GW) detection in the future, such as with the Laser Interferometer Space Antenna (LISA; \citealt{LISA2017arXiv170200786A}). UCXBs are one of the anticipated sources of mHz GWs \citep{Nelemans2009CQGra..26i4030N, Chen2020ApJ...900L...8C}, where mHz GW detectors may detect many systems that are currently not known from electromagnetic observations. Simulations indicate that $\mathcal{O}(5)$ UCXBs that evolved from the main sequence channel should be detectable as mHz GW sources \citep{Chen2025ApJ...981..175C}, as we suggest is the evolution scenario for \target. Using \texttt{LEGWORK} \citep{LEGWORK2022ApJS..260...52W}, \target\ will be detectable at $5\sigma$ with LISA in 4 years if the orbital period and donor star mass are, for example: $20$~min and $\gtrsim0.013$~\Msun; $30$~min and $\gtrsim0.05$~\Msun; or $60$~min and $\gtrsim0.3$~\Msun, respectively. Thus, it is possible that \target\ will be a detectable mHz GW source if its orbital period is sufficiently short; the donor mass requirements for the shorter orbital periods in particular are indeed reasonable and similar to estimates for known UCXBs \citep[e.g.,][]{Heinke2013ApJ...768..184H}. A stricter estimate of its strain will be possible when an orbital period is confirmed or better constrained for \target, at which time it may become a candidate verification binary for LISA and other mHz GW detectors.

\subsection{X-ray spectral and timing features} \label{sec:xray}

\subsubsection{X-ray spectrum}

The blackbody spectral component corresponds to a small emitting area of a $0.21\pm0.02$~km radius from the normalization of \texttt{bbodyrad}. This is much smaller than the size of a neutron star, and is also inconsistent with a black hole accretor. This suggests either a hotspot most likely located on the magnetic pole, or a thin boundary or spreading layer belt. The former scenario may indicate a somewhat strong magnetic field that channels the accreting matter onto the magnetic pole. Combining this with the lack of a spin period found in the timing analysis then suggests a scenario where the obliquity angle between the spin and magnetic axes is small such that the hotspot flux remains largely unmodulated by the spin of the NS. The latter scenario is a small boundary or spreading layer that produces the thermal emission. The lack of a hotspot spectral component then may indicate that the magnetic field is weak as in most LMXBs, and thus that the neutron star likely has a $\sim$milliseconds spin period. 

The X-ray spectrum overall is very consistent with typical hard state LMXBs. NSXBs with very strong magnetic fields show flatter power laws than we find here in \target\ \citep{White1983ApJ...270..711W}, thus spectrally it is unlikely that there is a very strong magnetic field of $\gtrsim10^{12}$~G, independent of the lack of detected pulsations.

The weak iron line detected in the NICER spectrum with an equivalent with of 8\,eV is a notable feature. There are a few reasons why there may be a weak iron line in an XRB: (1) a highly ionized disk; (2) a disk that is truncated far from the accretor; and (3) high abundances of carbon or oxygen that screen the iron fluorescence \citep{Koliopanos2013MNRAS.432.1264K}. We can safely rule out a highly ionized disk given the low X-ray luminosity of \target. A truncated accretion disk is possible and rather expected given the low accretion rate, and that we do not find a reasonable accretion disk component to be present in the X-ray spectrum. High carbon or oxygen abundances are also possible if the system is a UCXB with a carbon- or oxygen-rich donor. Given that the X-ray spectrum is not reasonably fit with a \texttt{diskbb}, we can attribute the weak iron line feature to a truncated accretion disk, which has been observed in other NSXBs at low luminosities \citep[e.g.,][]{Degenaar2017MNRAS.464..398D, vandenEijnden2020MNRAS.493.1318V}. This reinforces that \target\ is accreting at a very low rate which is only possible at a very close distance given its X-ray flux. A carbon- or oxygen-rich donor may also be feasible in addition to a truncated accretion disk.

Spectral investigations of UCXBs have found both strong \citep[e.g.,][notably at a low $L_X$ of $\sim10^{35}$~\ergps]{Degenaar2017MNRAS.464..398D} and undetected \citep[e.g.,][]{Campana2003ApJ...594L..39C, Miller2003ApJ...583L..99M, Sanna2018A&A...610L...2S} iron lines, as well as long-term variable iron line strengths \citep{Koliopanos2021MNRAS.500.5603K}, uncorrelated with luminosity or spectral state. Therefore, the weak iron line here in \target\ is inconclusive for the UCXB population especially given the inferred truncated disk above.

\subsubsection{X-ray timing properties}

First, we further discuss the nondetection of a pulse period of $<1\%$, which has some relevance to the NS magnetic field. Many NS LMXBs do not have known spin periods, so this is not unusual. Pulsations are likely due to moderately strong magnetic fields ($B\gtrsim10^{9}$~G) that are able to channel accreting material and create hotspots on the NS surface \citep{Bildsten1997ApJS..113..367B, Patruno2021ASSL..461..143P}. Furthermore, of the systems that do show pulsations, some have only sporadically detected epochs of pulsations \citep[e.g.,][]{Altamirano2008ApJ...674L..45A, Casella2008ApJ...674L..41C}. The reasons governing why pulsations appear in some systems during certain times is not clear. Thus, we cannot rule out a moderately strong magnetic field similar to accreting millisecond X-ray pulsars \citep{Wijnands1998Natur.394..344W}, though a very strong field ($B\geq10^{12}$~G) is highly unlikely.

\begin{figure*}[ht!]
\sidecaption
\includegraphics[width=0.3\textwidth]{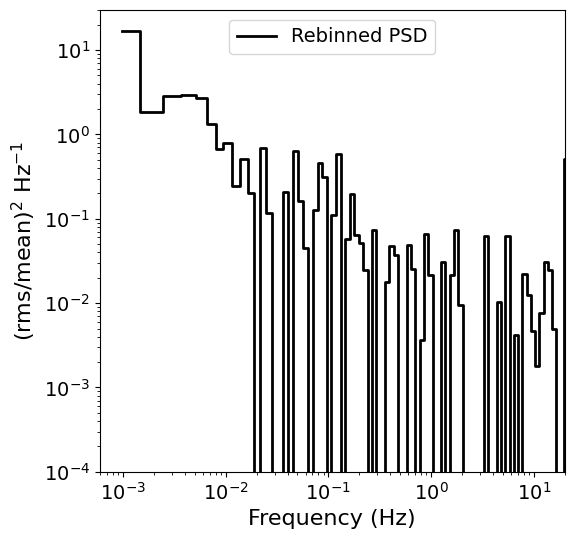} \hspace{0.2cm}
\includegraphics[width=0.3\textwidth]{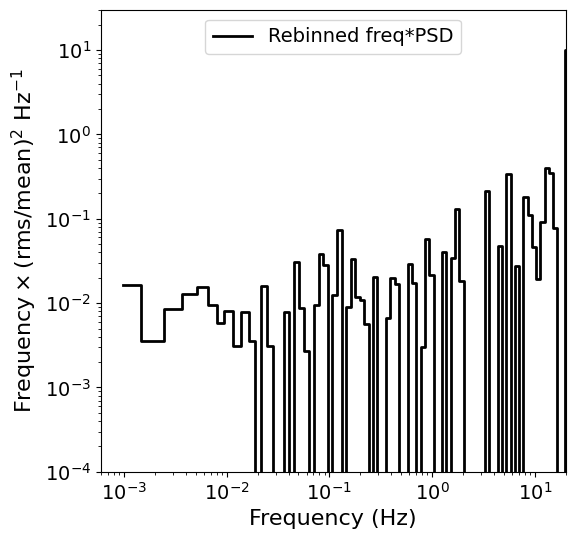}
    \captionsetup[figure]{justification=centering}
  \caption{Averaged power spectrum of \target\ of NICER data from \texttt{stingray}. On the left, the PSD shows a very low frequency noise component that remains unbroken to at least near $10^{-3}$~Hz. This is a very low frequency constraint on the break frequency value compared to other XRBs, an indication of a low intrinsic X-ray luminosity. On the right, the fractional power of $10^{-2}$ is also consistent with low accretion rate hard states in other XRBs, and is much higher than during soft states.}
     \label{fig:VLFN}
\end{figure*}

Other X-ray timing features of \target\ point toward a very low accretion rate. The fractional RMS of the NICER data is $25$\%, which is comparable to observations of hard states and is much higher than the low RMS of a few percent during soft states \citep{MunozDarias2011MNRAS.410..679M}.
The turnover frequency of the power spectrum of the very low frequency noise (VLFN) component is observationally positively correlated to the X-ray luminosity, and thus accretion rate \citep[e.g.,][]{Reig2004ApJ...602..918R, ArmasPadilla2014MNRAS.439.3908A}. We find that there is no turnover observed in \target\ down to $\sim$$10^{-3}$~Hz (Fig.~\ref{fig:VLFN}, left). The NSXB Aql~X--1 had an observed turnover at a frequency of 0.1~Hz at an X-ray luminosity of $\sim$$7\times10^{35}$~\ergps\ \citep{Reig2004ApJ...602..918R}, and the very faint BHXB transient Swift J1357.2-0933 had an observed turnover at 0.01~Hz at an X-ray luminosity of $\sim$$10^{35}$~\ergps\ \citep{ArmasPadilla2014MNRAS.439.3908A}. The fractional power of both systems was about $10^{-2}$, consistent with the fractional power observed in \target\ (Fig.~\ref{fig:VLFN}, right). The lack of an observed turnover in the PSD suggests that \target's X-ray luminosity is likely similar to or lower than the observations of the systems above.

\subsection{Radio jet} \label{sec:radio_jet}

A correlation between the X-ray and radio luminosities of XRBs is observed, particularly well in BHXBs \citep{Fender2004, Gallo2018MNRAS.478L.132G} (Fig.~\ref{fig:lrlx}), indicative of disk-jet coupling behavior and generally fit with a power law slope parameter $\beta$. BHXBs follow a close correlation across a wide range of X-ray luminosities, with $\beta \simeq 0.7$ \citep{Gallo2006, Gallo2018MNRAS.478L.132G}. However, it is less clear to date how NSXBs are correlated as there is a much wider range of radio luminosities for these systems. There have indeed been different $\beta$ values estimated for different classes of NSXBs, as well as among different NSXBs in the same class, ranging from $\simeq 0.4$ to $\simeq 1.5$ \citep{Qiao2019MNRAS.487.1626Q}, suggesting additional disk-jet complexity compared to BHXBs. The radio luminosity normalization of NSXBs is certainly lower on average than BHXBs (by a factor of $\sim22$; \citealt{Gallo2018MNRAS.478L.132G}) and spans at least two orders of magnitude across observed NSXBs to date \citep{vandenEijnden2021}, and may depend on factors such as the NS's intrinsic magnetic field strength and its spin \citep[e.g.,][]{vandenEijnden2018}.

As discussed before, the upper limit on \target's radio luminosity of $\sim 3 \times 10^{25}$ \ergps\ strongly indicates that the accretor is a NS rather than a BH, as this value is about 3 orders of magnitude fainter than the BHXB correlation at a similar X-ray luminosity of $10^{34}$~\ergps. There are also numerous radio jet detections of BHXBs at similar and lower X-ray luminosities as seen in Fig.~\ref{fig:lrlx}. Thus, a jet is expected to be present in \target\ due to its hard spectral state, and be easily detectable by our high-sensitivity radio observation if this system contained a BH accretor. 

Our radio nondetection leaves two general possibilities for the jet of \target\ with the assumption that it is a NSXB. One is that there is indeed a jet being produced that is very weak and thus undetected, through a steep $\beta$ slope, a low radio normalization factor (see Fig.~\ref{fig:lrlx}), or a combination of both. The other is that there is no jet being produced in this low accretion rate NSXB. Some BHXBs have confirmed radio emission during quiescence, attributed to a radio jet \citep{Gallo2014MNRAS.445..290G}; however, no quiescent ($L_X \simeq 10^{32}$~\ergps) NSXB has been detected in radio, thus the existence of radio jets in any class of NSXBs down to very low accretion rates remains unclear. It may be possible that at low accretion rates, the NS magnetosphere is able to quench the jet production in some way; that is, that there is a $\beta$ correlation above some $L_X$ and a steep drop below it, but to date this is unknown.  

It is also notable that there are indeed NSXB radio detections at higher radio luminosities for similar X-ray luminosities as \target. These are either accreting millisecond X-ray pulsars (AMXPs; \citealt{Wijnands1998Natur.394..344W}), or transitional millisecond pulsars (tMSPs; \citealt{Archibald2009Sci...324.1411A}). AMXPs have higher magnetic field strengths than the general NS LMXB population as evidenced by their X-ray pulsations, and tMSPs are a rare and less understood type of NSXB where the radio emission during their XRB phase does not behave as expected of a canonical compact jet \citep{Bogdanov2018ApJ...856...54B} (perhaps due to a propeller regime; \citealt{Illarionov1975A&A....39..185I}). There is no evidence that \target\ is a tMSP \citep{Papitto2022ASSL..465..157P}, as it is a persistent XRB without the characteristic step-function-shaped X-ray light curve of tMSPs, and is neither a radio nor gamma-ray source.
The radio jets of AMXPs are puzzling, as they host some of the brightest and faintest NSXB radio jets relative to their X-ray luminosities \citep{Tetarenko2016MNRAS.460..345T, Tudor2017MNRAS.470..324T}, thus a comparison of \target's radio jet luminosity limit to AMXP observations at similar X-ray luminosities is also not revealing.

A compilation of radio emission from a number of UCXBs by \citet{Tetarenko2018ApJ...854..125T} and recently expanded with new observations by \citet{Dage2025ApJ...988..131D} do not find a correlation between orbital period and radio luminosity within the UCXB population. The radio luminosities of UCXBs are consistent with the longer period population of LMXBs as well. Thus, to date there is no correlation between radio luminosity and orbital period of XRBs. This may indeed be expected as the jets are theoretically launched by magnetic fields in the vicinity of the accretor \citep{Blandford1977MNRAS.179..433B, Blandford1982MNRAS.199..883B}, which are not correlated with the size of the accretion disk, though there may also be some effect on jet production due to the altered mass:charge ratio of hydrogen-poor accretion disk material feeding the jet. Therefore, we are unable to draw a conclusion about \target's orbital period based on its radio nondetection.

Interestingly, there has been a NS UCXB observed to display drastic radio luminosity fluctuations, X1850--087 \citep{Panurach2023ApJ...946...88P}. This is a persistent system with $L_X\simeq10^{36}$~\ergps, and its radio jet was observed to vary significantly by a factor of $>10$ despite only small changes in the X-ray luminosity over several days. The cause of this radio variability is not known definitively, but it was suggested that there may be strong jet quenching below some $L_X$ value. This is similar to one of the scenarios we have considered for the very low $L_R$ upper limit of \target. Though, there is no indication that this jet variability is related to a UCXB nature, as there may also be some similarities with longer period AMXPs discussed above.

The very low radio luminosity limit of \target\ due to its close distance has provided a strong constraint in new \LRLX\ parameter space on the behavior of some NSXB jets at low X-ray luminosities, and motivates further deep radio monitoring of similar systems to better fill in the luminosity gap between \target\ and the majority of the radio-detected NSXB population at higher X-ray luminosities.

\subsection{Considering a longer orbital period or larger distance}

Here we consider whether the orbital period of \target\ might not be ultracompact and instead may be a more common LMXB value of a few hours, and whether it may be located at a distance much greater than $700$~pc.

\subsubsection{A longer orbital period at 700~pc}

The immediate difficulty with an orbital period longer than $\sim2$ hours is the faintness of the OIR counterpart, and the infrared one in particular as discussed in Section~\ref{sec:OIR_luminosity}. A longer orbital period would necessitate that the donor star be significantly underluminous while filling its Roche lobe, which is unreasonable. Therefore, if \target\ is indeed located at its \Gaia\ parallax distance, its orbital period cannot be longer than about 2 hours, and it should indeed be shorter than this when considering that the disk contributes significantly to the OIR emission.

\subsubsection{At a Galactic Center distance}

Another possibility is that the \Gaia\ parallax is either not accurate, or is affected by an interloper star such that the distance to \target\ is significantly larger than $700$~pc. These are unlikely given the $>5\sigma$ parallax, the consistency of the proper motion, and the extremely low probability of either dwarf stars or white dwarfs superimposed on the \Gaia\ counterpart. However, if \target\ is indeed located at a larger distance, the optical luminosity will increase, which allows for a nominally brighter and larger companion star and thus longer orbital period based on the OIR luminosity alone.

At a larger distance of, for example, $7$~kpc, the radio nondetection is notable again. \target\ would be one of the lowest radio upper limits for an XRB with $L_R\lesssim3\times10^{27}$~\ergps\ at similar X-ray luminosities of $L_X\simeq10^{36}$~\ergps\ (Fig.~\ref{fig:lrlx}). It would again lie in a parameter space that is outside of the observed range of \LRLX\ for BHXBs (this includes the radio-faint BHXB track at similar X-ray luminosities; \citealt{Coriat2011MNRAS.414..677C}), indicating a NS accretor regardless of its distance.

The X-ray luminosity at $7$~kpc becomes a more commonly observed persistent XRB value of $\sim10^{36}$~\ergps. The X-ray spectral and timing features remain the same; that is, the accretion rate must be consistent with a hard spectral state, which is reasonable if $L_X\simeq10^{36}$~\ergps\ or $1\%$~\Ledd. The thermal X-ray spectral component is still not compatible with a disk due to the inner radius remaining much smaller than the size of a neutron star (disk inner radius from the normalization parameter of up to 2~km at $7$~kpc). The vPM94 orbital period estimation is also affected, where the orbital period increases with distance, for example, to a nominal value of $1.1\pm0.9$~hours at $7$~kpc ($6\pm5$ hours at $+0.5~\mathrm{log}~\Sigma$). At an orbital period of a few hours, hydrogen is indeed expected in the spectrum, and the inclination angle estimations are the same.

There is indeed a concern though, as an increased distance and thus X-ray luminosity also raises the issue of the burst rate again. During hard states at $10^{36}$~\ergps, NS LMXBs routinely burst every few hours or so \citep[e.g.,][]{Cornelisse2003A&A...405.1033C, Galloway2020ApJS..249...32G} as the elevated mass accretion rate onto the NS quickly builds up a layer of hydrogen or helium to critical mass and temperature for ignition. 
As discussed above, there are no indications that \target\ has a very strong magnetic field that would inhibit bursting behavior, and it being an LMXB generally disfavors a highly magnetized scenario as well due to the presumed recycling of the NS accretor \citep{Bhattacharya1991PhR...203....1B, Wijnands1998Natur.394..344W} (though there is also a known highly magnetized NS UCXB, the $7.7$~s pulsar 4U~1626--67; \citealt{Middleditch1981ApJ...244.1001M, Orlandini1998ApJ...500L.163O}). However, due to the lack of evidence that \target\ hosts a strong magnetic field, we assume that it does not and thus there is no reason that it would not be frequently bursting at a luminosity of $10^{36}$~\ergps\ at a distance of several kpc with a helium- and possibly hydrogen-rich donor. This is a serious issue which strongly indicates that \target\ cannot be at a larger distance of several kpc, regardless of its orbital period at that distance.

\subsubsection{On the far side of the Galaxy}

A final consideration is if \target\ might be much further away than the Galactic Center, and may possibly be a halo object. There are indeed a couple of LMXBs with distances estimated to be $>20$~kpc \citep{Casares2004ApJ...613L.133C, Homan2011ATel.3650....1H}. At a distance of $20$~kpc, $L_X \simeq 10^{37}$~\ergps\ or $10\%~\mathrm{L_{Edd}}$, which is on the border between hard and soft states in XRBs \citep[e.g.,][]{Barret1996ApJ...473..963B}. A soft accretion state would significantly ease the tension with the burst rate or the radio nondetection; however, as discussed above, the timing (strong variability) and spectral (nonthermal-dominated) properties of the X-ray data clearly indicate a hard state. Thus, requiring that \target\ be in a hard state effectively limits its distance to $\lesssim20$~kpc. Then again, in a hard state similar to the $7$~kpc case above, there is a severe issue with the lack of thermonuclear bursts for a NSXB actively accreting at a relatively high rate with a presumed hydrogen- and helium-rich donor star.

\section{Summary}

The \Gaia\ parallax distance in combination with old and new data have enabled a fresh look at the parameters of \target, and a very faint ultracompact NSXB at $\sim$700\,pc interpretation is able to explain many of its observed characteristics. We summarize our various findings below.

\subsection{Basic properties}

    \begin{itemize}

    \item \Gaia\ has determined the distance of the optical counterpart of \target\ to be $\sim$700\,pc with no evidence of, and a very low probability for, an interloper star. This is one of the closest X-ray binaries known to date, and is the closest known Roche lobe overflowing LMXB.
  
    \item The X-ray luminosity is stably around $2 \times 10^{34}$~\ergps, an unusually low luminosity for a persistent system, or alternatively an unusually bright quiescence. MAXI and targeted X-ray data indicate that there has been no significant long-term X-ray variability of \target\ since its discovery more than 50 years ago, including no outbursts.
  
    \item New radio data have provided a deep upper limit for \target's radio jet, and it is about 3 orders of magnitude fainter than the BHXB correlation in \LRLX\ and well below the scatter of the BHXBs. This almost certainly determines that \target\ hosts a neutron star accretor. 

    \end{itemize}

\subsection{Binary orbit and companion or composition}

    \begin{itemize}
  
    \item The optical and infrared luminosities are extremely low, thus the orbital period must be very short, and is likely to be ultracompact. The X-ray and optical luminosities following \citet{vanParadijs1994A&A...290..133V}, as well as with the infrared counterpart following \citet{Russell2006MNRAS.371.1334R}, broadly indicate a very short orbital period of a few to tens of minutes. The infrared counterpart alone also conservatively restricts the orbital period to a maximum of $\sim2$~hours.

    \item There have been hydrogen emission disk lines observed in the optical spectrum, though the strength varied significantly in two separate observations. A UCXB may have residual hydrogen if the donor is an evolved star, which is a rare evolutionary channel for UCXBs.

    \item Using the H$\alpha$ FWHM-K2 relation of \citet{Casares2015ApJ...808...80C}, we determine that \target\ must be oriented at a very low inclination angle for any reasonable orbital period. This would also explain the lack of an observed orbital periodicity in the NICER data.

\end{itemize}

\subsection{X-ray results}

\begin{itemize}
  
    \item The X-ray spectrum is fit with a power law and blackbody component. A disk blackbody results in an unphysical inner disk radius, thus the spectrum is consistent with a truncated accretion disk, also aligning with the low accretion rate and potentially the effect of a NS magnetosphere. The small blackbody component is also inconsistent with a black hole accretor.
  
    \item There is a weak iron line with an equivalent width of $\sim$8\,eV. This can be explained by a truncated accretion disk, in line with the spectroscopic results.
  
    \item An additional explanation for both the weak iron line and lack of bursts is that the accretion disk may have significant abundances of carbon or oxygen. This would be expected if \target\ has a very short orbital period and thus a carbon- or oxygen-rich donor star, as suggested above, though the low X-ray luminosity is sufficient to explain both of these findings regardless of the composition of the disk.
  
    \item There is no indication of an orbital ($5\sigma < 2-5\%$) or spin period ($5\sigma < 1\%$) in recent X-ray data. The lack of an orbital period can be explained by a low inclination angle, and the lack of a spin period suggests either that the NS accretor has a weak magnetic field as is common in NS LMXBs, or that the spin and magnetic axes are closely aligned.
  
    \item The X-ray power spectrum is dominated by a very low frequency noise component commonly observed in hard state XRBs, and the break frequency is constrained to be $\lesssim 10^{-3}$~Hz. This is a very low value, which suggests that the X-ray luminosity is intrinsically very low, and is lower than typical hard state NSXB observations at $\sim10^{36}$~\ergps.
  
    \item No Type~I bursts have been observed from \target\ despite its NS nature, which can also now be explained by its very low accretion rate with predicted recurrence times on the order of months to years. If it does burst, it may be a rare and bright intermediate duration burst.

\end{itemize}

\subsection{Radio and other results}

\begin{itemize}

    \item The radio jet is undetected with an upper limit of $5\times10^{25}$~\ergps, one of the lowest radio upper limits obtained for an XRB, with two general possibilities. One is that a jet is present but much weaker than some AMXPs and tMSPs at similar X-ray luminosities. We rule out \target\ being a tMSP, but are unable to draw a conclusion about similarities with AMXPs. The other is that at this low accretion rate, there is NS magnetospheric jet quenching regardless of the $\beta$ slope or normalization parameter in \LRLX, which is not observed in BHXBs. This may be a vital clue as to the ability of NSXBs to launch jets at very low accretion rates.
  
    \item A persistently active XRB at a low X-ray luminosity is more easily sustained in a shorter orbital period system than a longer one, as it is easier to keep a physically small accretion disk ionized, thus also mildly supporting an ultracompact nature. Alternatively, a persistent low, but higher than quiescent, X-ray luminosity has been observed in other UCXBs, possibly due to an elevated mass transfer rate from a degenerate donor due to gravitational wave radiation. 

    \item It may be possible that \target\ is at a much larger distance if the \Gaia\ parallax is incorrect. In this case the radio nondetection still indicates a NS accretor, with an upper limit of $\sim20$~kpc for it to remain spectrally in a hard state. However, the lack of bursting at a higher X-ray luminosity with a helium-rich and presumed hydrogen-rich donor is not able to be explained, and a black hole accretor is still unreasonable. Placing \target\ at a larger distance seems more problematic than at 700~pc.
  
    \item A UCXB at 700\,pc is a promising future mHz gravitational wave source if its orbital period is sufficiently short. A future detection of \target's orbital period will allow for a more accurate estimate of its strain.

\end{itemize}

\target\ has become an intriguing XRB in light of recent data, and we have formed a novel interpretation that explains many previous uncertainties regarding this system and its properties. We have determined that \target\ is located at a distance of only $\sim$700\,pc from its \Gaia\ parallax, the closest low-mass Roche lobe overflowing XRB known to date. It is a NSXB existing at an unusual persistent X-ray luminosity of $2\times10^{34}$~\ergps, though it has never been observed to burst which is now explainable by its very low accretion rate. It is very likely to be an ultracompact system due to its high X-ray:optical flux ratio and overall faint OIR counterpart, though still with residual hydrogen in its disk. It hosts a very weak or absent radio jet, for which we have obtained an upper limit at one of the lowest radio luminosities ever observed in an XRB, and in an unprecedented parameter space on \LRLX. Its close proximity will be beneficial to multiwavelength follow-up in the near future to better refine various parameters, and particularly to identify its orbital period to confirm the theory of \target\ presented here.

\begin{acknowledgements}

We thank the anonymous referee for providing useful suggestions which improved the manuscript.
This research has made use of data and/or software provided by the High Energy Astrophysics Science Archive Research Center (HEASARC), which is a service of the Astrophysics Science Division at NASA/GSFC. The Australia Telescope Compact Array is part of the Australia Telescope National Facility (grid.421683.a) which is funded by the Australian Government for operation as a National Facility managed by CSIRO. ECP and TJM acknowledge funding via the NICER Guest Observer program, through grant 80NSSC23K1119. This research was supported by Deutsche Forschungsgemeinschaft  (DFG, German Research Foundation) under Germany’s Excellence Strategy - EXC 2121 "Quantum Universe" – 390833306. Co-funded by the European Union (ERC, CompactBINARIES, 101078773). Views and opinions expressed are however those of the author(s) only and do not necessarily reflect those of the European Union or the European Research Council. Neither the European Union nor the granting authority can be held responsible for them. We acknowledge the Gomeroi people as the traditional owners of the ATCA observatory site. We thank Jakob van den Eijnden, Duncan K. Galloway, James C. A. Miller-Jones, and Anna L. Watts for useful discussions. 

\end{acknowledgements}

\bibliographystyle{aa}
\bibliography{Bib}{}

\end{document}